\documentclass[10pt,letterpaper]{article}

\usepackage{cogsci}

\cogscifinalcopy 
\usepackage{graphicx}
\usepackage[export]{adjustbox}
\usepackage{pslatex}
\usepackage{apacite}

\usepackage{float} 
\usepackage{soul,color}
\usepackage{subcaption}
\usepackage{color}
\usepackage[usenames,dvipsnames]{xcolor}

\title{Computational characterization of the role of an attention schema\\ in controlling visuospatial attention}
 
\author{{\large \bf Lotta Piefke (lotta.piefke@gmx.com)} \\
  Institute of Cognitive Science, Osnabrück University, Osnabrück, Germany
  \AND {\large \bf Adrien Doerig (adoerig@uos.de)} \\
  Institute of Cognitive Science, Osnabrück University, Osnabrück, Germany
  \AND {\large \bf Tim Kietzmann (tkietzma@uos.de)} \\
  Institute of Cognitive Science, Osnabrück University, Osnabrück, Germany
  \AND {\large \bf Sushrut Thorat (sthorat@uos.de)} \\
  Institute of Cognitive Science, Osnabrück University, Osnabrück, Germany}

\begin{document}

\maketitle

\begin{abstract}

How does the brain control attention? The Attention Schema Theory suggests that the brain explicitly models its state of attention, termed an attention schema, for its control. However, it remains unclear under which circumstances an attention schema is computationally useful, and whether it can emerge in a learning system without hard-wiring. To address these questions, we trained a reinforcement learning agent with attention to track and catch a ball in a noisy environment. Crucially, the agent had additional resources that it could freely use. We asked under which conditions these additional resources develop an attention schema to track attention. We found that the more uncertain the agent was about the location of its attentional window, the more it benefited from these additional resources, which developed an attention schema. Together, these results indicate that an attention schema emerges in simple learning systems where attention is important and difficult to track\footnote{Training/analysis scripts to reproduce our results can be found at: https://github.com/KietzmannLab/Attention-Schema-Analysis}.

\end{abstract}

\section{Introduction}

Attention is a process through which the brain selects, and preferentially processes, parts of externally-driven or internally-generated information relevant to its behavior~\cite{hommel2019no}. In the visual domain, one type of attention - spatial attention - is often operationalized as a spotlight within which stimuli are processed with an enhanced signal relative to noise~\cite{posner1980orienting,eriksen1986visual}. How is the deployment of such attention controlled? Systems such as the fronto-parietal network and the superior colliculus are involved in attentional control~\cite{szczepanski2010mechanisms,corbetta2011spatial,krauzlis2013superior}. However, the computations underlying attentional control are poorly understood~\cite{petersen2012attention}.

Inspired by model-based control theory~\cite{conant1970every} and primate motor control studies~\cite{graziano2002brain,holmes2004body}, one account - the Attention Schema Theory (AST) - proposes that the brain controls its attention by building a descriptive and predictive model of attention (termed ``attention schema") ~\cite{graziano2015attention,graziano2011human}. This theory is supported by human studies showing that when the model of attention is disrupted, endogenous control of attention is affected~\cite{webb2016effects}. Normative modeling approaches have provided preliminary evidence that such an ``attention schema" aids in attentional control~\cite{van2017neurologically,wilterson2021attention,liu2023attention}. To test the usefulness of the schema in attention control, in \citeA{wilterson2021attention} a reinforcement learning agent was trained to track a ball in a noisy environment and move a paddle to catch it (see Figure~\ref{figure1}a). They compared agents with or without an attention schema. This attention schema was hard-wired to explicitly encode the attentional state, i.e., the spatial location and extent of the attention spotlight. They found that the schema was essential in learning attentional control. However, it was unclear (a) if the system required a hard-wired attention schema to track attention or if it could learn such a schema, and (b) what aspects of the environment and the agent made the schema essential.

To answer these questions, we conducted experiments in a setting adapted from \citeA{wilterson2021attention}, allowing the agent to access additional resources decoupled from the input that it could learn to use as it wished. Importantly, we also decoupled the additional resources from the attentional state. This is a crucial departure from \citeA{wilterson2021attention}, who hard-wired these additional resources to track the attentional state.  We found that the agent autonomously learned to use the additional resources to track its attentional state better. As the stimulus already provided some information about the attentional state, the schema did not need to be a copy of the attentional state but only to provide further hints about the location of the attentional state. Relatedly, we found that the usefulness of the schema was proportional to how poorly the attentional state can be inferred solely from the stimulus, given the noise in the environment. In sum, we characterized when an attention schema emerges (when it is essential in attentional control) and the information therein (hints about the attentional state).

\section{Methods}

\subsection{The agent and its environment}

\begin{figure}[!h]
\centering
\includegraphics[scale=0.65,center]{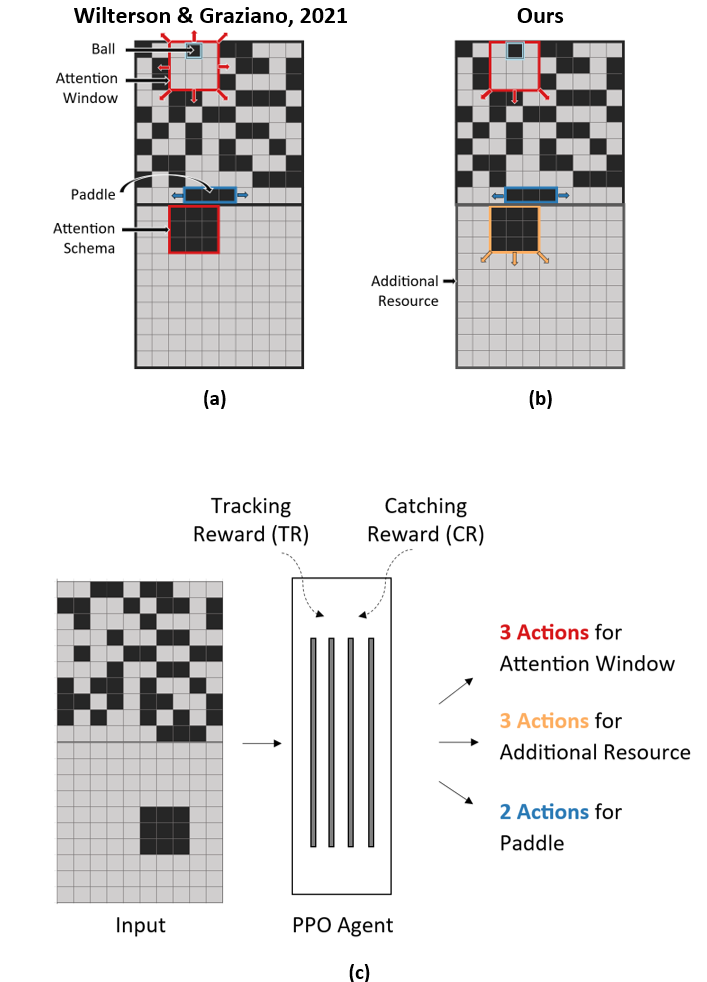}
\caption{The agent and its environment. Agents were trained to track a ball (highlighted here in teal) with an attention window (highlighted here in red), that removed noise inside it, and catch it with a paddle (highlighted here in blue). It could avail of an additional controllable resource to perform its task. (a) In Wilterson and Graziano
(2021), the additional resource was coupled with the attention window i.e. it was hard-wired as an attention schema. (b) In our setup, the black square in the additional resource had its own set of learnable actions (highlighted here in orange), so we could assess if an attention schema was essential, if it emerged, and what information it carried. (c) The input-output details of the Proximal Policy Optimization (PPO) agent are shown. The $4$-layer actor-network is depicted.}
\label{figure1}
\end{figure}

We adapted the setup from \citeA{wilterson2021attention}. The environment is schematized in Figure~\ref{figure1}a: the starting (top row) and end (bottom row) column positions of the ball (which is a black pixel) are randomly chosen. The ball moves diagonally downwards until its column position aligns with the end position and then it moves vertically down. At each timestep, the grid is flooded with random noise - each pixel has a probability $p$ to be turned black, independently at each timestep. $p$ was set to 0.5 in the original study. An attention window is initialized around the ball at the start of the trial, inside which noise is absent, simulating visuospatial attention (referred to as ``attended stimulus"). The objective of the agent is to move its attention window such that it can keep track of the ball, and move the paddle at the bottom such that it can catch that ball. The agent can make use of an additional resource to keep track of its attentional state. 

To assess the emergence of an attention schema, we consider a simple deviation from the original setting. In \citeA{wilterson2021attention}, the attentional schema was hard-wired to be a copy of the attentional window. In our setup, we decoupled the attention window from the additional resource. We set the additional resource to contain a black square with the same extent as the attention window while allowing the agent to move the square, as seen in Figure~\ref{figure1}b. The square was initialized at the location of the attention window at the start of each trial. 

The agent is schematized in Figure~\ref{figure1}c. Instead of the Deep Q Network~\cite{mnih2013playing} used in \citeA{wilterson2021attention}, we used the more recent Proximal Policy Optimization (PPO) algorithm~\cite{schulman2017proximal}. In \citeA{wilterson2021attention}, the agent remembers the past by storing each input frame in a memory bank, but we removed the memory bank to reduce computational complexity. PPO is an actor-critic algorithm. Here, the actor-network has four layers, and the critic network has three layers, with $1000$ units per layer. The agent has to perform three kinds of actions: 1) move the paddle, 2) move its attention window, and 3) decide what to do with its additional neural resources. In \citeA{wilterson2021attention}, the attention window has $8$ possible actions, as seen in Figure~\ref{figure1}a. As our setup required the introduction of actions for the additional resources, to make learning easier for the agent, we restricted the attention window and additional resources action spaces to $3$ actions each: down, down-left, and down-right, as seen in Figure~\ref{figure1}b (this makes sense, since the ball always moves down at every timestep). At each timestep, if the ball was within the attention window, the agent accrued a $0.5$ ball-tracking reward (TR), and if it was not then a penalty of $0.5$ was accrued. If the paddle caught the ball at the end of the trial, the agent received a ball-catching reward (CR) of $2$, else it received a penalty of $2$\footnote{Preliminary experiments revealed that the task is learnable without the ball tracking reward but it takes much longer to achieve good performance - after $80\,$k epochs, CR $\sim 1.25$}. 

All agents were trained until convergence of TR+CR near the optimum (max $=6$) i.e. the mean reward of the last $200$ test epochs was greater than $5.95$\footnote{The number of epochs for the agents trained (in environments with different noise probabilities) is as follows: $379$ $(p=0)$, $11486$ $(p=0.25)$, $5797$ $(p=0.5)$, $5090$ $(p=0.75)$, $222$ $(p=1)$}. For our analysis, each agent's behavior was observed over $2000$ new trials with random start/end positions for the ball and random noise.

\subsection{Analysis}

We assessed whether an attention schema emerges in the additional resources. In other words, we tested if the agent learned to use the additional resource to track the location of its attention. The additional resource is defined to contain an attention schema if it contains information about the attentional state, analogous to the definition of a body schema~\cite{de2021body}. Additionally, if such a schema needs to be a copy of the attentional state, we reasoned the agent could learn a 1-to-1 mapping between the location of the square and the attention window. 

One agent was trained per noise probability (Figure~\ref{figure3}a). For all the agents and further manipulations (e.g. ablation of the additional resource), the mean accrued rewards, on the test set, are reported with $95\%$ bootstrap confidence intervals of the mean over $10000$ samples, sampled with repetition. 

To diagnose how much information the agent could contain about a variable (e.g. the attentional state), we trained networks with the same architecture as the actor-network (which is used by the agent in its behavior) to infer the variable from the desired input (e.g. solely from the attended stimulus). In the classification analysis, where inference about the attentional state is made using various inputs, the mean of the cross-validation ($10$-fold) accuracies are reported.

\section{Results}

\subsection{Attention schema emerges for attentional control}

\citeA{wilterson2021attention} showed that their hard-wired attention schema was essential in controlling attention and performing the task. We assessed a) if our agent also relied on its additional resource to perform its task, and b) if it used the additional resource to track attention. For noise probability $p=0.5$, we observed that our trained agent accumulated a mean ball-tracking reward (TR) of $3.74[3.70, 3.79]$ (max = $4$) and a ball-catching reward (CR) of $1.73[1.68, 1.77]$ (max = $2$). When we removed the additional resource, the mean TR dropped to $-0.82[-0.93,-0.70]$, and the mean CR dropped to $-0.89[-0.97,-0.81]$. A better, i.i.d., control is randomizing the actions of the resource, as the agent was never trained to work without the additional resource. Randomizing the actions of the resource also led to the mean TR dropping to $0.93[0.81, 1.05]$, and the mean CR dropping to $-0.03[-0.12, 0.06]$. Furthermore, training the agent without the additional resource\footnote{The training did not converge according to our criterion. We report the performance after training the agent for a similar number of epochs as the agent trained with the additional resource.} led to a mean TR of $-0.39[-0.52, -0.26]$ and a mean CR of $-0.84[-0.92, -0.76]$. These results suggest that the agent relied on the additional resources to perform the task. Without the additional resources, the agent could not properly track or learn to track the ball to catch it.

How did the agent use the attentional resource? The Attention Schema Theory suggests an attention schema is essential. Thus, we assessed if this useful additional resource carried explicit information about the attentional state i.e. is it an attention schema? We trained a network with the same architecture as the actor-network to infer the location of the center ($64$ locations; $8$ rows $\times$ $8$ columns) of the attention window from the additional resource, across timesteps and trials. The attentional state could be inferred given the additional resource (Test accuracy: $61\%$, chance: $1.6\%$, randomized actions: $40\%$). We also assessed if this inference was solely driven by the row correspondence between the attention window and the black square in the additional resource, as they both move down one pixel per timestep. The column of the attentional state could also be inferred given the additional resource (Test accuracy: $60\%$, chance: $12.5\%$, randomized actions: $40\%$). This result suggests that, in learning to use its additional resources for performing its task, the agent acquired an attention schema.

Is this schema actually useful in allowing the agent to infer its attentional state - does it provide more information about the attentional state than the attended stimulus does? To answer these questions, we trained a network with the same architecture as the actor-network to infer the location of the center of the attention window from either the attended stimulus alone or the full input (the attended stimulus and the additional resource), across timesteps and trials. The attentional state could be inferred better with the inclusion of the additional resource (stimulus - 60\%, stimulus + resource - 81\%, chance - 1.6\%). Thus, the agent can improve knowledge of its attentional state by using the attention schema in the additional resource.

\begin{figure}[h]
\centering
\begin{subfigure}[b]{0.5\textwidth}
    \includegraphics[scale=0.8,center]{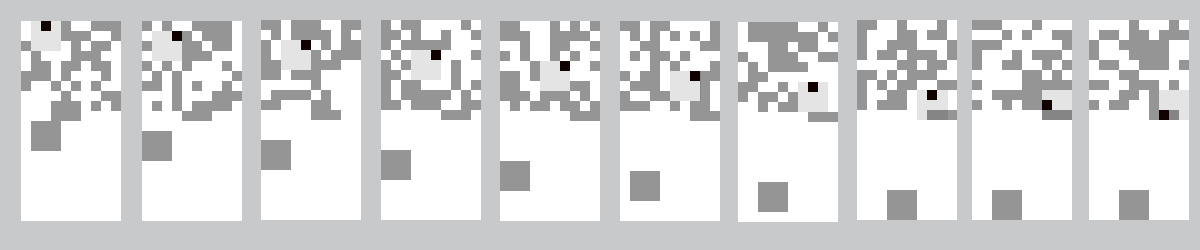}
    \caption{}
\end{subfigure}
\begin{subfigure}[b]{0.5\textwidth}
    \includegraphics[scale=0.55]{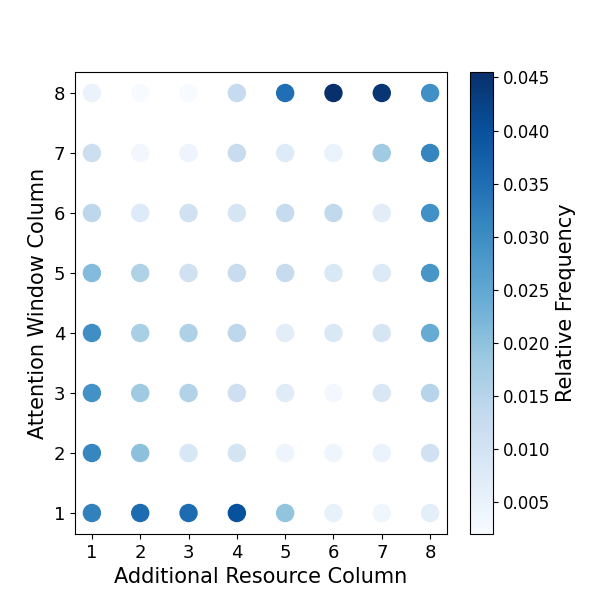}
    \caption{}
\end{subfigure}
\caption{Characterizing the emergent attention schema. (a) An example trial of the trained agent. The square in the additional resource does not precisely track the attention window (highlighted here as the light grey box) that contains the ball (highlighted here as the black pixel), although it is essential for directing the attention window. (b) There is no 1-to-1 correspondence between the column of the black square of the additional resource and the attention window. Instead, the additional resource only provides hints about the location of the attention window.} 
\label{figure2}
\end{figure}

Given that the additional resource resembles an attentional schema, we asked whether the learned attention schema contains an explicit copy of the attentional state, similar to the hard-wiring by \citeA{wilterson2021attention}. Visualizing the agent's behavior (Figure~\ref{figure2}a) indicated that the black square in the additional resource did not move exactly like the attention window did, i.e., the attention schema was not an exact copy of the attention location. To quantify this observation, we assessed whether the attention schema had a 1-to-1 correspondence with the location of the attention window. As the rows would always correspond since both the attention window and the black square move down one pixel per timestep, we focused on the columns. In Figure~\ref{figure2}b, we plot the relative frequency of observing the centers of the black square in the additional resource and the attention window across columns, across timesteps and trials. We did not observe a 1-to-1 correspondence. This dovetails with our finding that although the attentional state can be inferred from the additional resources, the inference is not perfect. In sum, the emergent attention schema does not have a 1-to-1 correspondence with the attentional state. Instead, it contains information that can provide hints to the agent to improve the attentional state inference.

Given that the additional resource contained a schema that did not perfectly signal information about the attentional state we asked whether this was the case due to the agent's performance being worse than if the schema was hard-wired. An agent for which the additional resource contained a hard-wired attention schema, as in \citeA{wilterson2021attention}, was trained. It accumulated a mean ball-tracking reward (TR) of $3.73[3.68, 3.78]$ and a ball-catching reward (CR) of $1.77[1.726, 1.808]$, which was not significantly higher than our agent (TR: $3.74[3.70, 3.79]$, CR: $1.73[1.68, 1.77]$). These results suggest that an explicit copy of the attentional state is not essential for attentional control - hints can suffice.

\begin{figure*}[!h]
\centering
\begin{subfigure}[b]{\textwidth}
    \includegraphics[scale=0.3, center]{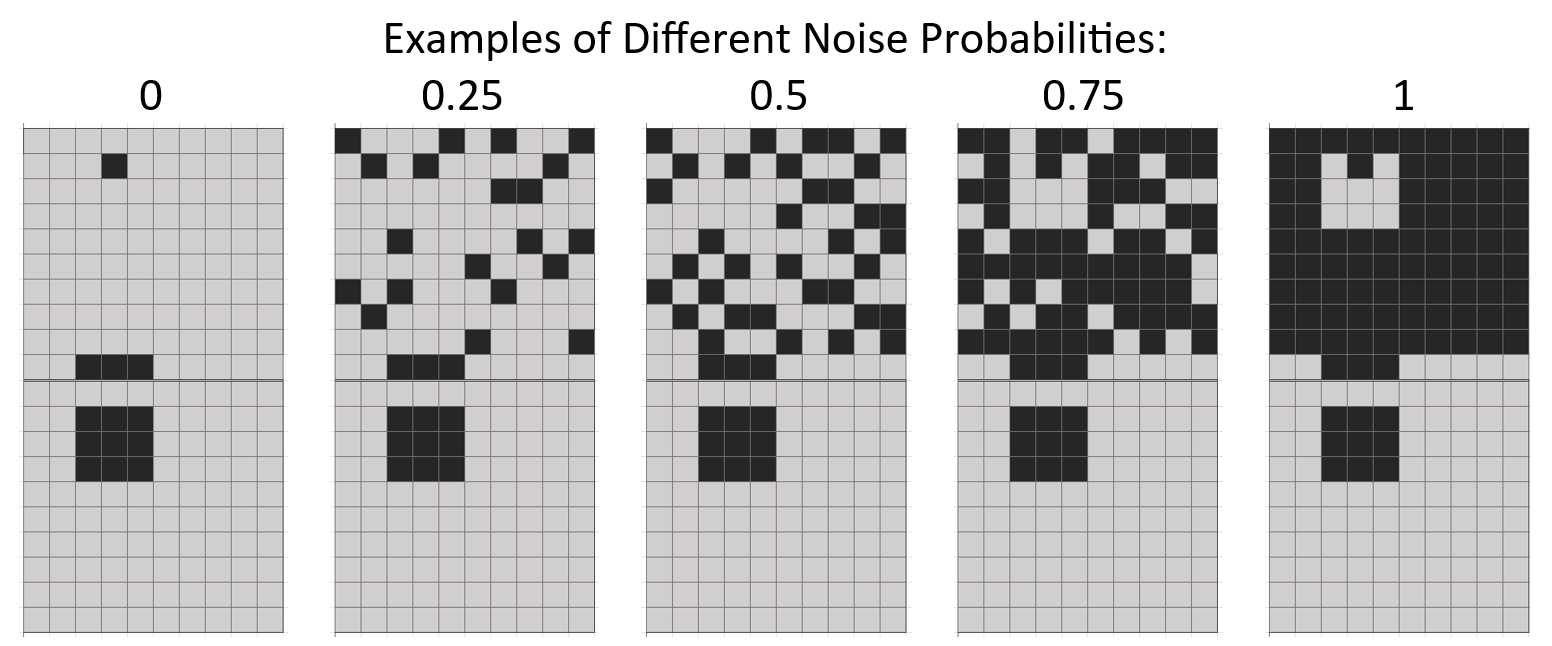}
    \caption{}
\end{subfigure}
\begin{subfigure}[b]{0.5\textwidth}
    \includegraphics[scale=0.55]{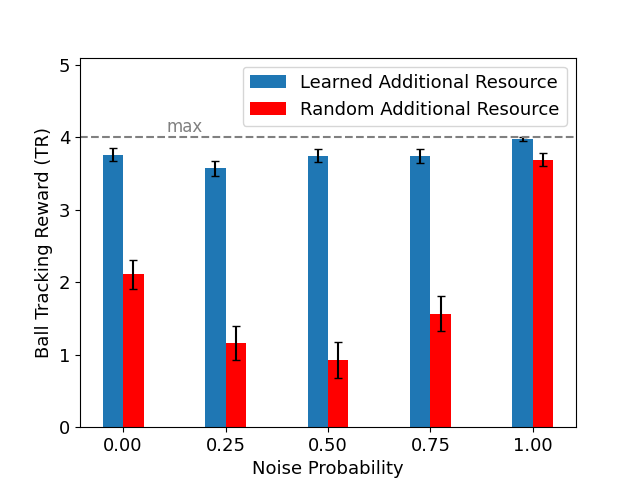}
    \caption{}
\end{subfigure}\hfill
\begin{subfigure}[b]{0.5\textwidth}
    \includegraphics[scale=0.55]{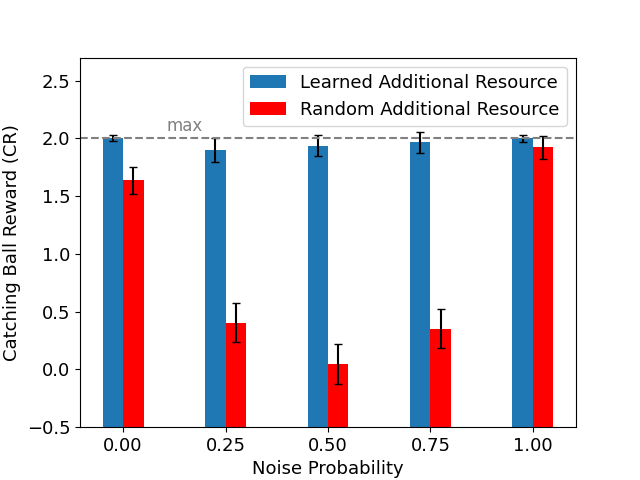}
    \caption{}
\end{subfigure}

\begin{subfigure}[b]{0.5\textwidth}
    \includegraphics[scale=0.55]{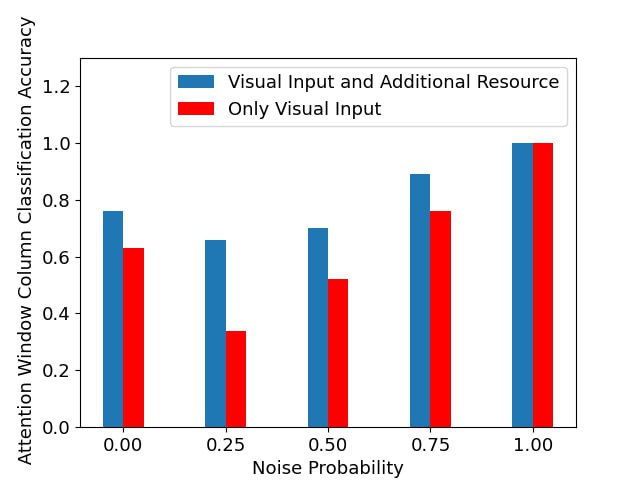}
    \caption{}
\end{subfigure}\hfill
\begin{subfigure}[b]{0.5\textwidth}
    \includegraphics[scale=0.55]{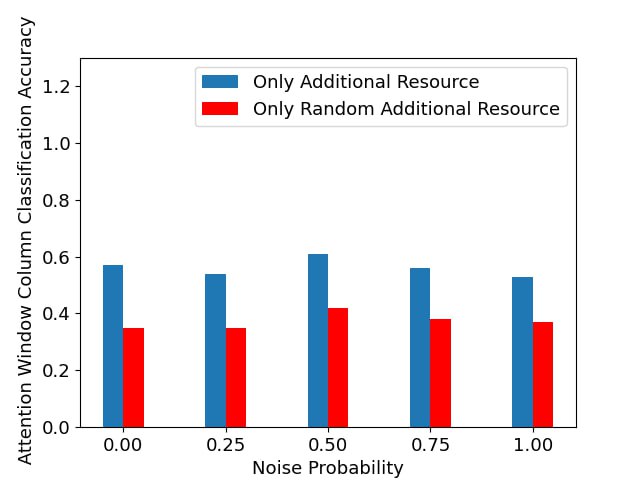}
    \caption{}
\end{subfigure}

\caption{Usefulness of the attentional schema as a function of the noise probability. (a) Examples of the environment containing varying noise probabilities are shown. With the additional resource, the agent can acquire (b) near-perfect attention tracking (max. reward = $4$) and (c) near-perfect ball tracking (max. reward = $2$). When the actions of the additional resource are randomized, the agent suffers the most in intermediate noise probabilities i.e. the additional resource is most useful in environments with intermediate noise probabilities. (d) Under intermediate noise probabilities, the additional resource adds the most information to the stimulus in inferring the attentional state, as shown by the difference in the accuracies in classifying the position of the attention window given the stimulus alone or the stimulus and the additional resource. (e) Under intermediate noise probabilities, the additional resource also provides the most information about the attentional state, as shown by the accuracies in classifying the position of the attention window given the additional resource alone.} 
\label{figure3}
\end{figure*}

\subsection{Attention schema is useful when the attentional state cannot be inferred from the stimulus}
 
In the previous section, we saw that the attention schema was essential in attentional control when the environment had a noise probability of $0.5$. We now assess whether the usefulness of the attention schema is dependent on the noise in the environment. To quantify the usefulness of the schema, we monitor the ball-tracking reward (TR) and the catching reward (CR). Usefulness is measured as the reward deficit created when the learned actions for the additional resource are randomized: if the randomization reduces TR and CR, then the schema is useful. We assessed the accrued rewards as a function of the noise probability in the environment, where the agent was trained, and whether the learned additional resource actions were left intact or randomized. As seen in Figure~\ref{figure3}b-c, the usefulness of the attention schema was maximal for $0.5$ noise probability. The deficit in TR and CR caused by randomizing the additional resource actions (quantifying the usefulness of the schema) is reduced with lower or higher noise probabilities. 


To understand why the usefulness of the schema was maximal for intermediate noise probabilities, we asked if the usefulness was related to how easy the attentional state inference was solely given the noise probability in the environment. For example, when $p=1$, the entire environment is filled with black pixels (see Figure~\ref{figure3}a), except the attention window, which makes the attentional state inference trivial, and possibly reduces the need for an attention schema. We trained a network with the same architecture as the actor-network to predict the column position of the attention window from the stimulus alone (no additional resource provided as input), or to predict the same given the stimulus and the additional resource. We focussed on the column position as the row position always increases by $1$ with each timestep for both the additional resource and the attention window, which would positively bias the inference. The difficulty of attentional state inference is measured as the deficit created by omitting the additional resource. We assessed how the difficulty of attentional state inference varied with noise. As seen in Figure~\ref{figure3}d, the difficulty was maximal for a noise probability of $0.25$ and decreased for higher or lower noise. Qualitatively comparing Figure~\ref{figure3}d with Figure~\ref{figure3}b-c suggested that the deficit in attentional state inference due to the absence of the attention schema co-varied with the usefulness of the attention schema, across noise probabilities.

Next, we tested whether the agent developed a better attention schema when it was useful in performing the task. We trained a network with the same architecture as the actor-network to predict the column position of the center of the attention window from the additional resource alone (with the learned actions or randomized actions). We assessed how the inference of attentional state varied with noise. First, as seen in Figure~\ref{figure3}e, the information about the attention window was higher given the learned additional resource actions than given the randomized actions. This implied that the correspondence between the additional resource and the attentional state was higher due to an emergent schema. Second, the inference was best for a noise probability of $0.5$ and decreased slightly for higher or lower noise probabilities. Qualitatively, this suggests that the increase in correspondence between the additional resource and the attentional state is higher for noise probabilities where the additional resource is also useful (intermediate noise probabilities). In other words, the schema seems to contain better information about the attentional state when the schema is more useful for the task. However, it is interesting to note that the attentional state could be inferred better from the learned additional resource even when the noise probability was $1$, while randomization of the additional resource actions did not substantially reduce the performance of the agent. This indicates the additional resource acquires an attention schema even if it is marginally essential.

Together, our results suggest that an attention schema naturally emerges and aids attentional control in cases where the attentional state cannot be inferred solely from the attended stimulus due to noise.

\section{Discussion}

Previous work ~\cite{wilterson2021attention} showed that agents controlling visuospatial attention to track and catch a ball in noisy environments require an attention schema. In this study, we replicated and extended these findings with a modified agent. We find that an attention schema does not need to be hard-wired. It can emerge through learning, given additional resources. Moreover, it does not need to be a copy of the attentional state but only to provide hints that improve attentional state inference. We also found that the attention schema is more useful when the inference of the attentional state solely from the attended stimulus is harder. 

The primary deviation from \citeA{wilterson2021attention} in the current study was the decoupling of the additional resource actions from the attention window actions. After training on the tracking and catching task in a noisy environment, the additional resource contained information about the attentional state i.e. an attention schema existed in the additional resource. One limitation is that we still applied several constraints to the additional resource that the agent used to learn the attention schema (namely, the additional resource still contained a single 3x3 square, and only 3 actions were allowed to move it). Future research will endow the agent with an unstructured resource to check whether an attention schema emerges in this unconstrained computational resource, too.

The usefulness of the attention schema in attentional control was found to be higher when the environment contained intermediate noise probabilities. In such cases, the attentional state could not be inferred well solely from the input, which made the hints provided by the schema essential. The ability of the agent to infer the attentional state from the input might improve with larger networks, or if the agent has access to an explicit memory, as it did in \citeA{wilterson2021attention} (alternatively, such a memory might also emerge in an actor-network if the network is recurrent, allowing for the integration of information across time). Such dependence of the usefulness of the attention schema on factors such as network size and memory should be studied to further characterize the need for an attention schema.

In the current setting and in \citeA{wilterson2021attention}, the noise outside the attention window is not affected by the attention mechanism. However, in the brain, visuospatial attention leads to suppression of activity outside the attentional spotlight~\cite{posner1980orienting,desimone1995neural}. If such a mechanism is implemented in the current environment, potentially the inference of the attentional state given solely the attended stimulus would be easier, making the attention schema less important. However, in naturalistic settings, objects present outside the attentional spotlight can be salient enough to drive neural activity leading to distraction, which can be counteracted with sustained attention~\cite{kim2010neural,demeter2016transient}. In such naturalistic settings, inference of the attentional state might again be harder solely given the attended stimulus and might require an attention schema.

In summary, in this study, the additional resource contains a schema that provides partial information about the attentional state. This is sufficient for the agent as the attended stimulus also provides some information about the attentional state. It is a priori unclear why the attention schema needs to be a perfect copy of the attentional state, unless attentional state inference, as required by the task, is not possible at all from the attended stimulus. This notion that the attention schema does not need to be a copy of the attentional state raises questions about the content of awareness in primates. The Attention Schema Theory~\cite{graziano2015attention} proposes that the content of awareness draws upon the information contained in the attention schema. If the schema only contains partial information about the attentional state, perhaps that is also what we can be aware of during the task. Studying the content of the attention schema in agents operating in naturalistic environments might provide more specific Attention Schema Theory hypotheses for the content of awareness. This would allow for better tests of the validity of the Attention Schema Theory as a theory of consciousness.

\section{Acknowledgments}

The project was funded by the European Union (ERC, TIME, Project $101039524$). Compute resources were funded by the DFG (Project number $456666331$).

\bibliographystyle{apacite}

\setlength{\bibleftmargin}{.125in}
\setlength{\bibindent}{-\bibleftmargin}

\bibliography{lib}

\end{document}